\documentclass[prd,nofootinbib,superscriptaddress]{revtex4}
\usepackage[latin9]{inputenc}
\usepackage{amsmath, amssymb, graphicx, color, textcomp, amsfonts, graphics, epstopdf, slashed, multirow, adjustbox, lipsum, rotating, xcolor, wrapfig, epsfig}
\usepackage[unicode=true, bookmarks=false, breaklinks=false,pdfborder={0 0 1},backref=section,colorlinks=false]{hyperref}
\hypersetup{colorlinks,bookmarksopen,bookmarksnumbered,linkcolor=blu,pdfstartview=FitH,urlcolor=rossos,citecolor=rossos}
\usepackage[compat=1.1.0]{tikz-feynman}
\usepackage[justification=raggedright, margin=5mm,font=sf,small,labelfont=bf, format=plain]{caption}
\definecolor{rosso}{cmyk}{0,1,1,0.4}
\definecolor{rossos}{cmyk}{0,1,1,0.55}
\definecolor{rossoc}{cmyk}{0,1,1,0.2}
\definecolor{blu}{cmyk}{1,1,0,0.3}
\definecolor{blus}{cmyk}{1,1,0,0.6}
\definecolor{bluc}{cmyk}{1,1,0,0.1}
\definecolor{verde}{cmyk}{0.92,0,0.59,0.25}
\definecolor{verdec}{cmyk}{0.92,0,0.59,0.15}
\definecolor{verdes}{cmyk}{0.92,0,0.59,0.4}
\DeclareRobustCommand{\orcidicon}{\hspace{-2.1mm}
\begin{tikzpicture}
\draw[lime,fill=lime] (0,0.0) circle [radius=0.13] node[white] {{\fontfamily{qag}\selectfont \tiny ID}}; \draw[white,fill=white] (-0.0525,0.095) circle [radius=0.007];
\end{tikzpicture} \hspace{-3.7mm} }
\foreach \x in {A, ..., Z}
 {\expandafter\xdef\csname orcid\x\endcsname{\noexpand\href{https://orcid.org/\csname orcidauthor\x\endcsname} {\noexpand\orcidicon}}}

\begin{document}
\title{The 95 GeV Excess in the Georgi-Machacek Model: Single or Twin Peak
Resonance}
\author{Amine Ahriche\orcidA}
\email{ahriche@sharjah.ac.ae}
\affiliation{Department of Applied Physics and Astronomy, University of Sharjah, P.O. Box 27272 Sharjah, UAE.}
\begin{abstract}
In this work, we investigate the possibility to address the excess
observed around 95 GeV in the $\gamma\gamma$, $\tau\tau$, and $b\bar{b}$
channels as a scalar resonance(s) within the Georgi-Machacek (GM)
model. In our analysis, we find that the excess can be easily accommodated
in the channels ($\gamma\gamma$ and $b\bar{b}$) simultaneously,
where the 95 GeV candidate is a single peak resonance (SPR) due to
a light CP-even scalar. We found that the excess in the $\tau\tau$
channel can be addressed simultaneously with $\gamma\gamma$ and $b\bar{b}$
only if the 95 GeV candidate is a twin peak resonance (TPR), i.e.,
another CP-odd scalar in addition to the CP-even scalar. We demonstrate
that the nature of the 95 GeV scalar resonance candidate (SPR or TPR)
can be probed via the properties of its di-$\tau$ decay.
\end{abstract}
\maketitle

\section{Introduction}

Since the Higgs boson discovery with a mass around $125~{\rm GeV}$~
\cite{ATLAS:2012yve}, the question about how the electroweak
symmetry breaking (EWSB) proceeded, remains open. It is not clear
yet whether the EWSB proceeded via a single Higgs as in the standard
model (SM) or via many scalars as in many SM extensions. Therefore,
the LHC physics program is devoting many searches and analyses to
the search for a di-Higgs signal and the search for additional scalar
resonances, whether they are heavier or lighter than the $125~{\rm GeV}$
Higgs, for example see~\cite{ATLAS:2020tlo,CMS:2018rmh}.

Although many searches for light scalar ($\eta$) have been performed
at LEP and the LHC (8+13 TeV), where an excess around 95 GeV has been
reported in the channels~\cite{ATLAS_talk,CMS:2022goy,Cao:2016uwt,Azatov:2012bz}
\begin{align}
\mu_{\gamma\gamma}^{\mathrm{exp}} & =\mu_{\gamma\gamma}^{\mathrm{ATLAS+CMS}}=\frac{\sigma^{{\rm exp}}\left(gg\to\eta\to\gamma\gamma\right)}{\sigma^{{\rm SM}}\left(gg\to h\to\gamma\gamma\right)}=0.27_{-0.09}^{+0.10},\nonumber \\
\mu_{\tau\tau}^{{\rm exp}} & =\mu_{\tau\tau}^{\mathrm{CMS}}=\frac{\sigma^{{\rm exp}}\left(gg\to\eta\to\tau\tau\right)}{\sigma^{{\rm SM}}\left(gg\to h\to\tau\tau\right)}=1.2\pm0.5,\nonumber \\
\mu_{bb}^{{\rm exp}} & =\mu_{bb}^{{\rm LEP}}=\frac{\sigma^{{\rm exp}}\left(e^{+}e^{-}\to Z\eta\to Zb\bar{b}\right)}{\sigma^{{\rm SM}}\left(e^{+}e^{-}\to Zh\to Zb\bar{b}\right)}=0.117\pm0.057,\label{eq:excess}
\end{align}
with local significance values $3.2\,\sigma$~\cite{Biekotter:2023oen},
 $2.2~\sigma$~\cite{CMS:2022goy} and $2.3~\sigma$~\cite{Barate:2003sz},
respectively.

Regarding the significant difference between the observed values in
(\ref{eq:excess}), addressing the excess in the three channels simultaneously
via a single particle is a very difficult, see, for example~\cite{Cacciapaglia:2016tlr,Crivellin:2017upt,Cao:2019ofo,Biekotter:2019kde,Cline:2019okt,Abdelalim:2020xfk,Heinemeyer:2021msz,Biekotter:2021qbc,Biekotter:2021ovi,Li:2022etb,Biekotter:2022abc,Benbrik:2022tlg,Iguro:2022dok,Biekotter:2022jyr,Benbrik:2022azi,Biekotter:2023jld,Azevedo:2023zkg,Ahriche:2023hho,Chen:2023bqr,Dev:2023kzu,Li:2023kbf,Bhattacharya:2023lmu,Coloretti:2023wng,Ashanujjaman:2023etj,Liu:2024cbr,Cao:2024axg,Kalinowski:2024uxe,Ellwanger:2024txc,Arcadi:2023smv,Arhrib:2024wjj,Benbrik:2024ptw,Ayazi:2024fmn,Ellwanger:2024vvs,Wang:2024bkg}.
From model building point of view, in order to explain the excess
in the different channels (\ref{eq:excess}) as a scalar resonance,
the $95~{\rm GeV}$ scalar candidate should have SM-like couplings.
In many new physics (NP) models, the $95~{\rm GeV}$ scalar resonance
candidate exhibits couplings to both gauge bosons and fermions that
scale similarly with respect to the SM values, i.e., $\frac{g_{\eta ff}^{NP}}{g_{\eta VV}^{NP}}\sim\frac{g_{hff}^{SM}}{g_{hVV}^{SM}}$.
A NP model can successfully address the excess in these channels simultaneously
if the couplings of the $95~{\rm GeV}$ scalar resonance candidate
to the SM fermions and gauge bosons are uncorrelated, i.e., $\frac{g_{\eta ff}^{NP}}{g_{\eta VV}^{NP}}\ne\frac{g_{hff}^{SM}}{g_{hVV}^{SM}}$.
This feature does exist in the so-called Georgi-Machacek (GM) model~\cite{Georgi:1985nv},
where it has been shown that a viable parameter space exists for the
light CP-even scalar case ($\eta$) with SM-like couplings~\cite{Ahriche:2022aoj}.
Many phenomenological aspects of this model have been extensively
studied in the literature~\cite{Chanowitz:1985ug,Gunion:1989ci,Haber:1999zh,Aoki:2007ah,Godfrey:2010qb,Low:2010jp,Logan:2010en,Chang:2012gn,Kanemura:2013mc,Englert:2013zpa,Killick:2013mya,Englert:2013wga,Ghosh:2019qie,Das:2018vkv,Hartling:2014zca,Hartling:2014aga,Chiang:2014bia,Chiang:2015rva,Chang:2017niy,Chiang:2015kka,Chen:2022zsh,Chen:2020ark,Pilkington:2017qam,Chiang:2014hia,Ismail:2020zoz,Bairi:2022adc,Ghosh:2023izq}.

The GM scalar sector has a residual global custodial $SU(2)_{V}$
symmetry after the EWSB, where its spectrum consists of two CP-even
singlets ($h$ and $\eta$), a triplet ($H_{3}^{0}$,$H_{3}^{\pm}$)
and a quintuplet ($H_{5}^{0}$,$H_{5}^{\pm}$,$H_{5}^{\pm\pm}$).
In this study, we explore whether the 95 GeV scalar resonance candidate
could be the CP-even scalar single peak resonance (SPR) $\eta$ (with
a mass around 95 GeV), or a twin peak resonance (TPR) consisting of
both the CP-even $\eta$ and CP-odd $H_{3}^{0}$, each with a degenerate
mass around $m_{\eta}\thickapprox m_{H_{3}^{0}}\sim95~{\rm GeV}$.
A viable parameter space can be defined by confronting these two possibilities
with the relevant theoretical and experimental constraints.

This work is organized as follows; in Section~\ref{sec:Model}, we
review the GM model where we define the mass spectrum and the relevant
scalar couplings. Then, in Section~\ref{sec:Constr}, we discuss
the different theoretical and experimental constraints relevant to
our study. We discuss the $95~{\rm GeV}$ signal excess in the $\gamma\gamma,~\tau\tau$
and $b\bar{b}$ channels within the GM model in Section~\ref{sec:excess};
where the relevant parameter space is identified. In Section~\ref{sec:Dis},
we discuss the possibility of distinguishing the SPR and TPR scenarios
using the di-$\tau$ channel. In Section~\ref{sec:Conclusion}, we
give our conclusion.

\section{Model, Mass Spectrum \& Couplings\label{sec:Model}}

In the GM model, the scalar sector consists of a doublet $(\phi^{+},\,\phi^{0})^{T}$;
a complex triplet $(\chi^{++},\,\chi^{+},\,\chi^{0})^{T}$ and a real
triplet $(\xi^{+},\,\xi^{0},\,-\xi^{-})^{T}$ with the hypercharge
$Y=1,2,0$, respectively;
\begin{equation}
\Phi=\left(\begin{array}{cc}
\frac{h_{\phi}-ia_{\phi}}{\sqrt{2}} & \phi^{+}\\
-\phi^{-} & \frac{h_{\phi}+ia_{\phi}}{\sqrt{2}}
\end{array}\right),\,\Delta=\left(\begin{array}{ccc}
\frac{h_{\chi}-ia_{\chi}}{\sqrt{2}} & \xi^{+} & \chi^{++}\\
-\chi^{-} & h_{\xi} & \chi^{+}\\
\chi^{--} & -\xi^{-} & \frac{h_{\chi}+ia_{\chi}}{\sqrt{2}}
\end{array}\right),\label{eq:field}
\end{equation}
where the tree-level custodial symmetry is ensured by the scalar vacuum
expectation values (VEVs) choice $\{\left\langle h_{\phi}\right\rangle ,\left\langle h_{\chi}\right\rangle ,\left\langle h_{\xi}\right\rangle \}=\{\upsilon_{\phi},\sqrt{2}\upsilon_{\xi},\upsilon_{\xi}\}$
with $\upsilon_{\phi}^{2}+8\upsilon_{\xi}^{2}\equiv\upsilon_{SM}^{2}$.
The GM scalar potential is invariant under the global symmetry $SU(2)_{L}\times SU(2)_{R}\times U(1)_{Y}$;
and given by
\begin{align}
V(\varPhi,\Delta) & =\frac{m_{1}^{2}}{2}\mathrm{Tr}[\varPhi^{\dagger}\varPhi]+\frac{m_{2}^{2}}{2}\mathrm{Tr}[\Delta^{\dagger}\Delta]+\lambda_{1}(\mathrm{Tr}[\varPhi^{\dagger}\varPhi])^{2}+\lambda_{2}\mathrm{Tr}[\varPhi^{\dagger}\varPhi]\mathrm{Tr}[\Delta^{\dagger}\Delta]+\lambda_{3}\mathrm{Tr}[(\Delta^{\dagger}\Delta)^{2}]+\lambda_{4}(\mathrm{Tr}[\Delta^{\dagger}\Delta])^{2}\nonumber \\
 & -\lambda_{5}\mathrm{Tr}[\varPhi^{\dagger}\frac{\sigma^{a}}{2}\varPhi\frac{\sigma^{b}}{2}]\mathrm{Tr}[\Delta^{\dagger}T^{a}\Delta T^{b}]-\mu_{1}\mathrm{Tr}[\varPhi^{\dagger}\frac{\sigma^{a}}{2}\varPhi\frac{\sigma^{b}}{2}](U\Delta U^{\dagger})_{ab}-\mu_{2}\mathrm{Tr}[\Delta^{\dagger}T^{a}\Delta T^{b}](U\Delta U^{\dagger})_{ab},\label{eq:V}
\end{align}
where $\sigma^{1,2,3}$ are the Pauli matrices and $T^{1,2,3}$ correspond
to the generators of the $SU(2)$ triplet representation and the matrix
$U$ is given in~\cite{Georgi:1985nv}. The GM scalar potential (\ref{eq:V})
is invariant under the global symmetry $SU(2)_{L}\times SU(2)_{R}\times U(1)_{Y}$
that is broken to a residual $SU(2)_{V}$ during the EWSB. The scalar
spectrum includes three CP-even eigenstates $\{h_{\phi},h_{\chi},h_{\xi}\}\to\{h,\eta,H_{5}^{0}\}$,
a neutral Goldstone and CP-odd eigenstate $\{a_{\phi},a_{\chi}\}\to\{G^{0},H_{3}^{0}\}$,
a charged Goldstone and two singly charged scalars $\{\phi^{\pm},\chi^{\pm},\xi^{\pm}\}\to\{G^{\pm},H_{3}^{\pm},H_{5}^{\pm}\}$,
and one doubly charged scalar $\chi^{\pm\pm}\equiv H_{5}^{\pm\pm}$,
that are defined as~\cite{Georgi:1985nv}
\begin{align}
h & =c_{\alpha}h_{\phi}-\frac{s_{\alpha}}{\sqrt{3}}(\sqrt{2}h_{\chi}+h_{\xi}),\,\eta=s_{\alpha}h_{\phi}+\frac{c_{\alpha}}{\sqrt{3}}(\sqrt{2}h_{\chi}+h_{\xi}),\,H_{5}^{0}=\sqrt{\frac{2}{3}}h_{\xi}-\sqrt{\frac{1}{3}}h_{\chi},\nonumber \\
H_{3}^{0} & =-s_{\beta}a_{\phi}+c_{\beta}a_{\chi},\,H_{3}^{\pm}=-s_{\beta}\phi^{\pm}+c_{\beta}\frac{1}{\sqrt{2}}(\chi^{\pm}+\xi^{\pm}),\,H_{5}^{\pm}=\frac{1}{\sqrt{2}}(\chi^{\pm}-\xi^{\pm}),\,H_{5}^{\pm\pm}=\chi^{\pm\pm},\label{eq:Eigen}
\end{align}
with $s_{x}=\sin x$, $c_{x}=\cos x$ ($x=\alpha,\beta$); and $t_{\beta}\equiv\tan\beta=\sqrt{8}\upsilon_{\xi}/\upsilon_{\phi}$
and $\tan2\alpha=2M_{12}^{2}/(M_{22}^{2}-M_{11}^{2})$, where $M^{2}$
is the scalar squared mass matrix in the basis $\{h_{\phi},\,\sqrt{\frac{2}{3}}h_{\chi}+\frac{1}{\sqrt{3}}h_{\xi}\}$.
One has to mention that the CP-even scalar $H_{5}^{0}=\sqrt{\frac{2}{3}}h_{\xi}-\frac{1}{\sqrt{3}}h_{\chi}$
does not couple to the SM fermions, and therefore cannot play any
role in explaining this anomaly.

In this setup, the CP-odd scalar $H_{3}^{0}$ couples to the SM fermions
but not to both gauge bosons, however, the CP-even scalar $\eta$
(95 GeV candidate) has SM-like couplings to the gauge fields and fermions.
Let us define the $SX\bar{X}$ coupling modifiers with respect to the
SM in the GM model with $S=h,\eta,H_{3}^{0}$ and $X=\mu,\tau,b,c,W,Z,\gamma,\tilde{g}$.
Lets us call $\varrho_{X}=\frac{g_{SX\bar{X}}^{GM}}{g_{hX\bar{X}}^{SM}}$
for $\varrho_{X}=\kappa_{\mathfrak{\mathrm{X}}},\zeta_{X},\vartheta_{X}$,
i.e.,
\begin{align}
\kappa_{\mathfrak{\mathrm{F}}} & =\frac{g_{hff}^{GM}}{g_{hff}^{SM}}=\frac{c_{\alpha}}{c_{\beta}},\,\kappa_{V}=\frac{g_{hVV}^{GM}}{g_{hVV}^{SM}}=c_{\alpha}c_{\beta}-\sqrt{\frac{8}{3}}s_{\alpha}s_{\beta},\nonumber \\
\zeta_{F} & =\frac{g_{\eta FF}^{GM}}{g_{hFF}^{SM}}=\frac{s_{\alpha}}{c_{\beta}},\,\zeta_{V}=\frac{g_{\eta VV}^{GM}}{g_{hVV}^{SM}}=s_{\alpha}c_{\beta}+\sqrt{\frac{8}{3}}c_{\alpha}s_{\beta},\nonumber \\
\vartheta_{F} & =\frac{g_{H_{3}^{0}FF}^{GM}}{g_{hFF}^{SM}}=-s_{\alpha},\,\vartheta_{V}=\frac{g_{H_{3}^{0}VV}^{GM}}{g_{hVV}^{SM}}=0.\label{eq:Scaling}
\end{align}
Here, we have $F=\mu,\tau,b,c$ and $V=W,Z$. The scalar gluon effective
vertices are mediated by the top/bottom quark loops, which implies
$\varrho_{\tilde{g}}=\varrho_{F}$, while the scalar photon effective
coupling modifiers are given by
\begin{align}
\varrho_{\gamma} & =\left|\frac{\varrho_{V}A_{1}^{\gamma\gamma}(4m_{W}^{2}/m_{S}^{2})+\varrho_{F}\frac{4}{3}A_{1/2}^{\gamma\gamma}(4m_{t}^{2}/m_{S}^{2})+\varrho_{F}\frac{1}{3}A_{1/2}^{\gamma\gamma}(4m_{b}^{2}/m_{S}^{2})+\frac{\upsilon}{2}\sum_{i}\frac{g_{SXX}^{GM}}{m_{X}^{2}}Q_{X}^{2}A_{0}^{\gamma\gamma}(4m_{X}^{2}/m_{S}^{2})}{A_{1}^{\gamma\gamma}(4m_{W}^{2}/m_{S}^{2})+\frac{4}{3}A_{1/2}^{\gamma\gamma}(4m_{t}^{2}/m_{S}^{2})+\frac{1}{3}A_{1/2}^{\gamma\gamma}(4m_{b}^{2}/m_{S}^{2})}\right|,\label{eq:eGam}
\end{align}
where $X=H_{3}^{\pm},\,H_{5}^{\pm},\,H_{5}^{\pm\pm}$ stands for all
charged scalars inside the loop diagrams, $Q_{X}$ is the electric
charge of the field $X$ in units of $|e|$, $g_{SXX}^{GM}$ are triple
couplings of the scalar $S=h,\eta,H_{3}^{0}$ to the charged scalars,
respectively; and the one-loop functions $A_{i}^{\gamma\gamma}$ are
given in~\cite{Djouadi:2005gi}.

One has to mention that in the GM model, there exists an invariance under the transformation $(\upsilon_{\xi},\mu_{1,2}) \to (-\upsilon_{\xi},-\mu_{1,2})$, which means $V(\Phi, \Delta ,\mu_{1,2})=V(\Phi, -\Delta ,-\mu_{1,2})$. The scalar mass matrix elements also remain invariant under this transformation. However, because the physical scalar eigenstates are mixtures of the components of the doublet and triplets, most of the physical triple and quartic scalar vertices are not invariant under $(\upsilon_{\xi},\mu_{1,2}) \to (-\upsilon_{\xi},-\mu_{1,2})$. This implies that any two benchmark points (BPs) with the same input parameters but with different signs of $(\pm t_{\beta}, \pm \mu_{1,2})$ are physically different. This can be easily seen in the couplings modifier (\ref{eq:Scaling}); thus, negative $t_{\beta}$ values should not be ignored in the numerical scan.

\section{Theoretical \& Experimental Constraints\label{sec:Constr}}

In our analysis, we consider many theoretical and experimental constraints
such as tree-level unitarity, boundness from below, Higgs measurements
(total decay width and coupling modifiers); and different negative
searches at LEP and the LHC. These constraints are detailed in~\cite{Bairi:2022adc,Ahriche:2022aoj}.
It has been shown that the GM scalar potential may acquire some minima
that violate the CP symmetry or the electric charge, that are deeper
than the electroweak vacuum $\{\upsilon_{\phi},\sqrt{2}\upsilon_{\xi},\upsilon_{\xi}\}$.
This fact excludes about 40\% of the parameter space that is usually
considered in the literature~\cite{Bairi:2022adc}.

Since the scalar resonance candidates $\eta$ and $H_{3}^{0}$ mass
is around $95~{\rm GeV}$, their decay is mainly to $\mu\mu,\tau\tau,bb,cc,\gamma\gamma,\tilde{g}\tilde{g}$.
For the Higgs and the $95~{\rm GeV}$ scalar resonance candidates
$S=h,\eta,H_{3}^{0}$, one writes
\begin{align}
\varGamma_{S}^{tot} & =\Gamma_{S}^{SM}\sum_{X=SM}\varrho_{X}^{2}\mathcal{B}^{SM}(S\to XX),\,\mathcal{B}(S\to XX)=\varrho_{X}^{2}\Big(\Gamma_{S}/\Gamma_{S}^{SM}\Big)^{-1},\label{eq:decay}
\end{align}
where the SM numerical values of $\Gamma_{S}^{SM}$ and $\mathcal{B}^{SM}(S\to XX)$
are given in~\cite{Higgs}. This allows the partial signal strength
modifier at the LHC for the scalar $S$ to be simplified within the
Narrow Width Approximation (NWA) as
\begin{align}
\mu_{XX}^{S} & =\frac{\sigma(pp\to S)\times\mathcal{B}(S\to XX)}{\sigma^{SM}(pp\to h)\times\mathcal{B}^{SM}(h\to XX)}\nonumber \\
 & =\kappa_{F}^{2}\varrho_{X}^{2}\Big(\Gamma_{S}/\Gamma_{S}^{SM}\Big)^{-1},\label{eq:muXX}
\end{align}
where $\Gamma_{S}$ and $\Gamma_{S}^{SM}$ are the scalar total decay
width and its SM values, i.e., $\Gamma_{S}^{SM}=\Gamma_{h}^{SM}(m_{h}\to m_{S})$.
One has to mention that in this setup, the channels $h\to\eta\eta,H_{3}H_{3},H_{5}H_{5}$
and $\eta\to WW,ZZ,H_{3}H_{3},H_{5}H_{5}$ ($H_{3}^{0}\to WW,H_{5}H_{5}$)
are kinematically forbidden due to $m_{\eta}\sim95~{\rm GeV}$ ($m_{H_{3}^{0}}\sim95~{\rm GeV}$).

Here, we consider experimental measurements of the Higgs total decay width ($\Gamma_{h}=4.6_{-2.5}^{+2.6}~{\rm MeV}$~\cite{ParticleDataGroup:2022pth}),
the electroweak precision tests; and the Higgs strength signal modifiers
$\mu_{XX}^{h}$ for $X=\mu,\tau,b,\gamma,W,Z$~\cite{ParticleDataGroup:2022pth}.
One has to mention that the Higgs strength modifiers $\mu_{XX}^{h}$
for $X=\mu,\tau,b,W,Z$ can be obtained from (\ref{eq:muXX}) by replacing
$\varrho_{X}$ by $\kappa_{X}$ in (\ref{eq:Scaling}), while for
$\mu_{\gamma\gamma}^{h}$, $\varrho_{X}$ should be replaced by $\kappa_{\gamma}$
in (\ref{eq:eGam}). In order to ensure simultaneous matching for
all previous observables, we define a $\chi^{2}$ function
\begin{equation}
\chi_{SM}^{2}=\sum_{\mathcal{O}=1}^{8}\chi_{\mathcal{O}}^{2}=\sum_{\mathcal{O}=1}^{8}\left(\frac{\mathcal{O}-\mathcal{O}^{\mathrm{exp}}}{\Delta\mathcal{O}^{\mathrm{exp}}}\right)^{2},\label{eq:chi2-SM}
\end{equation}
where the observables $\mathcal{O}$ denote (1) the Higgs total decay
width ($\Gamma_{h}$), the Higgs signal strength modifiers (from 2
to 7) ($\mu_{XX}^{h}$ for $X=\mu,\tau,b,\gamma,W,Z$) and (8) the
oblique parameter $\Delta S$. One notices that the contribution of $\mu_{\gamma\gamma}^{h}$ to (\ref{eq:chi2-SM})
is more important than those of $\mu_{\mu\mu,\tau\tau,bb,WW,ZZ}^{h}$ since it depends on the charged scalar masses and scalar couplings in addition to the mixing angles $\alpha$ and $\beta$~\cite{Bairi:2022adc}. In our analysis, we consider a precision
of 95\% C.L., i.e., $\chi_{SM}^{2}<12.59$ for eight variables. The experimental
values of the oblique parameter in (\ref{eq:chi2-SM}) are $S=0.06\pm0.10$~\cite{ParticleDataGroup:2022pth}.
One has to mention that the oblique parameter $S$ was estimated in~\cite{Hartling:2014aga},
while the oblique parameter $T$ cannot be estimated in the GM model
since the hypercharge interactions break the $SU(2)_{R}$ global symmetry
at one-loop level, yielding a divergent value for the $T$ parameter~\cite{Gunion:1990dt,Englert:2013zpa}.

Besides the above-mentioned constraints, others should be considered
like the negative searches for doubly-charged Higgs bosons in the
VBF channel $H_{5}^{++}\rightarrow W^{+}W^{+}$; and the Drell-Yan
production of a neutral Higgs boson $pp\rightarrow H_{5}^{0}(\gamma\gamma)H_{5}^{+}$;
which gives strong bounds on the parameter space~\cite{Ismail:2020zoz}.
It has been shown in~\cite{Ismail:2020zoz}, that the doubly-charged
Higgs bosons in the VBF channel leads to important constraints from
CMS on $s_{\beta}^{2}\times\mathcal{B}(H_{5}^{++}\rightarrow W^{+}W^{+})$~\cite{CMS:2017fhs}.
The negative search of the quintet in the di-photon channel $H_{5}^{0}\rightarrow\gamma\gamma$
is translated into bounds on the fiducial cross section times branching
ratio $\sigma_{fid}=\big(\sigma_{H_{5}^{0}H_{5}^{+}}\times\epsilon_{+}+\sigma_{H_{5}^{0}H_{5}^{-}}\times\epsilon_{-}\big)\mathcal{B}(H_{5}^{0}\rightarrow\gamma\gamma)$,
which is constrained by ATLAS at 8 TeV~\cite{ATLAS:2014jdv} and at
13 TeV~\cite{ATLAS:2017ayi}. To incorporate these constraints into our numerical analysis, we utilized the formulas for the decay rate and the cross section, as well as the efficiency values used in~\cite{Ismail:2020zoz}. Regarding the bounds from the null results in searches for $H_{5}^{++}\rightarrow W^{+}W^{+}$, the CMS analysis~\cite{CMS:2017fhs} only considered masses of $m_{5}>200,\mathrm{GeV}$. Therefore, we extrapolated the existing bounds down to $m_{5}>78,\mathrm{GeV}$.

At LEP, the negative searches for SM-like light scalars at low mass
range $m_{\eta}<100~\textrm{GeV}$ impose a significant bound on
the cross section of $e^{-}e^{+}\to~\eta~Z$~\cite{OPAL:2002ifx},
i.e., the factor $\zeta_{V}^{2}$. However, one notices that this
bound is easily satisfied for the mass values around $95.4\,\mathrm{GeV}$~\cite{Ahriche:2022aoj}.
Another search of the light SM-like scalar in the di-photon channel with
masses in the range $70-110~\mathrm{GeV}$ has been performed by CMS
at 8 $\mathrm{TeV}$ and 13 $\mathrm{TeV}$~\cite{CMS:2018cyk},
where upper bounds are established on the production cross section
$\sigma(pp\to\eta)\times\mathcal{B}(\eta\to\gamma\gamma)$ scaled
by its SM value, i.e., the factor $\zeta_{F}^{2}.\zeta_{\gamma}^{2}$,
where $\zeta_{X}$'s are defined in (\ref{eq:Scaling}) and (\ref{eq:eGam}).
Concerning the CMS bounds~\cite{CMS:2018cyk} on the production cross
section of the CP-odd scalar $\sigma(pp\to H_{3}^{0})\times\mathcal{B}(H_{3}^{0}\to\gamma\gamma)$,
the bounds are automatically fulfilled since $|\vartheta_{F}.\vartheta_{\gamma}|<|\zeta_{F}.\zeta_{\gamma}|$
for all the viable parameter space.

Since the charged triplet $H_{3}^{\pm}$ is partially coming from
the SM doublet as shown in (\ref{eq:Eigen}), it then couples the
up to the down quark in a similar way that the W gauge boson does. These
interactions lead to flavor violating processes such as the $b\to s$
transition ones, which depend only on the charged triplet mass $m_{3}$
and the mixing angle $\beta$. The current experimental value of the
$b\to s\gamma$ branching ratio, for a photon energy $E_{\gamma}>1.6~\textrm{GeV}$
is $\mathcal{B}(\overline{B}\to X_{s}\gamma)_{exp}=(3.55\pm0.24\pm0.09)\times10^{-4}$,
while the two SM predictions are $\mathcal{B}(\overline{B}\to X_{s}\gamma)_{SM}=(3.15\pm0.23)\times10^{-4}$~\cite{Misiak:2006zs}
and $\mathcal{B}(\overline{B}\to X_{s}\gamma)_{SM}=(2.98\pm0.26)\times10^{-4}$~\cite{Becher:2006pu}.
In our numerical scan, we will consider the severe among the bounds
on the $m_{3}$-$\upsilon_{\chi}$ plan shown in Fig. 1 in~\cite{Hartling:2014aga}.

\section{The Excess in the $\gamma\gamma,~\tau\tau$ and $b\bar{b}$ Channels\label{sec:excess}}

Here, we estimate the excess observed by both LEP and LHC around the
$95.4\,\mathrm{GeV}$ mass value in the channels $\gamma\gamma,~\tau\tau,\,b\bar{b}$,
where the signal resonance is assumed to be a CP-even scalar for $93~{\rm GeV}<m_{\eta}<97~{\rm GeV}$;
or a superposition of two resonances if $93~{\rm GeV}<m_{\eta},m_{H_{3}^{0}}<97~{\rm GeV}$.
Then, the $95~{\rm GeV}$ signal resonance signal strength modifiers
can be written in the NWA as
\begin{align}
\mu_{\gamma\gamma}^{(95)} & =\mu_{\gamma\gamma}^{(\eta)}+\mu_{\gamma\gamma}^{(H_{3}^{0})}\nonumber \\
 & =\zeta_{F}^{2}\zeta_{\gamma}^{2}\Big(\Gamma_{\eta}/\Gamma_{\eta}^{SM}\Big)^{-1}+\vartheta_{F}^{2}\vartheta_{\gamma}^{2}\Big(\Gamma_{H_{3}^{0}}/\Gamma_{H_{3}^{0}}^{SM}\Big)^{-1},\nonumber \\
\mu_{\tau\tau}^{(95)} & =\mu_{\tau\tau}^{(\eta)}+\mu_{\tau\tau}^{(H_{3}^{0})}\nonumber \\
 & =\zeta_{F}^{4}\Big(\Gamma_{\eta}/\Gamma_{\eta}^{SM}\Big)^{-1}+\vartheta_{F}^{4}\Big(\Gamma_{H_{3}^{0}}/\Gamma_{H_{3}^{0}}^{SM}\Big)^{-1},\nonumber \\
\mu_{b\bar{b}}^{(95)} & =\mu_{b\bar{b}}^{(\eta)}=\zeta_{V}^{2}\zeta_{F}^{2}\Big(\Gamma_{\eta}/\Gamma_{\eta}^{SM}\Big)^{-1}.\label{eq:mu95}
\end{align}

One remarks that the signal $\mu_{b\bar{b}}^{(95)}$ does not include
the contribution $\mu_{b\bar{b}}^{(H_{3}^{0})}$ since the CP-odd
scalar $H_{3}^{0}$ does not couple to the $Z$ gauge boson. Clearly,
the CP-even scalar $H_{5}^{0}$ cannot play a similar role as $H_{3}^{0}$
since it does not couple to quarks and therefore cannot be ggF produced
at the LHC; and it also does not decay into the SM fermions if produced
at LEP.

In order to estimate the relative contributions $\rho_{\gamma\gamma,\tau\tau}=\mu_{\gamma\gamma,\tau\tau}^{(H_{3}^{0})}/\mu_{\gamma\gamma,\tau\tau}^{(95)}$
in (\ref{eq:mu95}), one has to mention that because the $H_{3}^{0}$
total decay width is much smaller than its corresponding SM value,
then the factor $\big(\Gamma_{H_{3}^{0}}/\Gamma_{H_{3}^{0}}^{SM}\big)^{-1}$
may lead to a significant enhancement for $\mu_{\gamma\gamma,\tau\tau}^{(95)}$.
In addition, the effective coupling modifier $\vartheta_{\gamma}$ is
very suppressed due to the absence of the gauge and scalar contributions
to $\vartheta_{\gamma}$ in (\ref{eq:eGam}). This makes the ratio
$\rho_{\tau\tau}=\mu_{\tau\tau}^{(H_{3}^{0})}/\mu_{\tau\tau}^{(95)}$
comparable to unity; but $\rho_{\gamma\gamma}=\mu_{\gamma\gamma}^{(H_{3}^{0})}/\mu_{\gamma\gamma}^{(95)}$
is very suppressed, as will be shown next.

By considering all the above mentioned constraints discussed in Section~\ref{sec:Constr},
we perform a numerical scan, where the masses lie in the ranges $93~\textrm{GeV}<m_{\eta}<97~\textrm{GeV}$
and $78~\textrm{GeV}<m_{3},m_{5}<2~\textrm{TeV}$, where $m_{3,5}$
are the triplet and quintuplet masses, respectively. The triplet mass
ranges $93~\textrm{GeV}<m_{3}<97~\textrm{GeV}$ correspond to the
TPR case; while the rest of the $m_{3}$ values correspond to the SPR scenario.
In addition, we impose the SM-like Higgs constraints to be fulfilled
at 95\% C.L. by taking $\chi_{SM}^{2}<12.59$, where $\chi_{SM}^{2}$
is defined in (\ref{eq:chi2-SM}).

\begin{figure}[h]
\includegraphics[width=0.48\textwidth,height=0.3\textwidth]{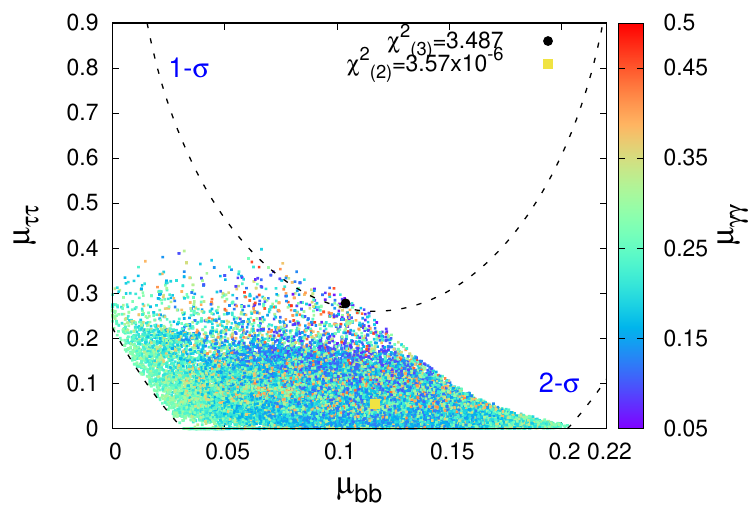}~\includegraphics[width=0.48\textwidth,height=0.3\textwidth]{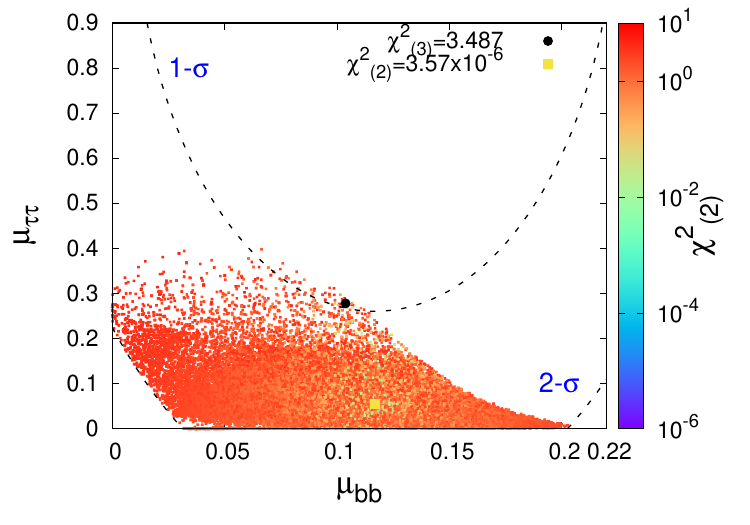}\\
 \includegraphics[width=0.48\textwidth,height=0.3\textwidth]{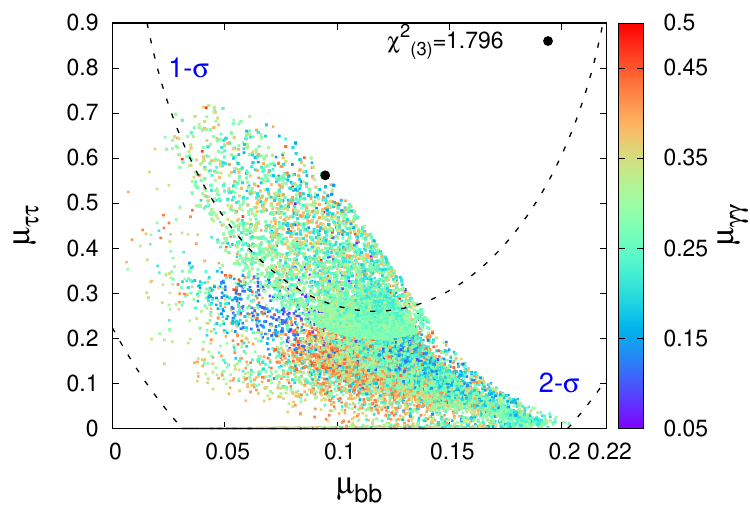}~\includegraphics[width=0.48\textwidth,height=0.3\textwidth]{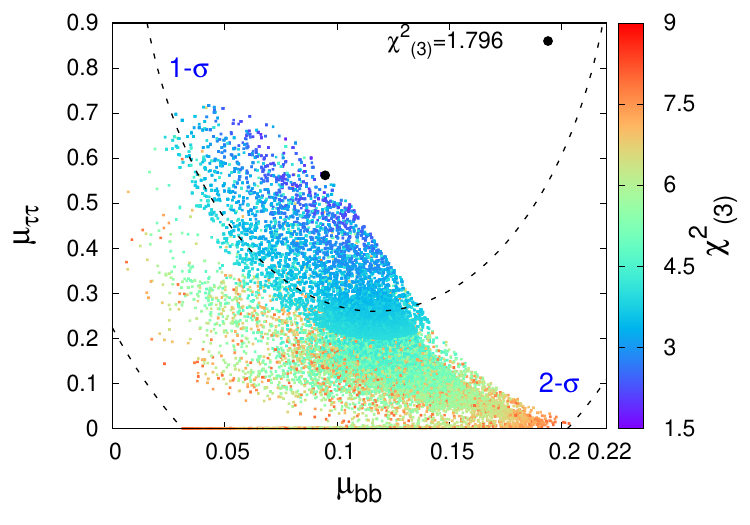}\caption{The signal strength values (\ref{eq:mu95}) for $2\sigma$ viable
BPs with $\chi_{(3)}^{2}<8.02,\,\chi_{SM}^{2}<12.59$ in the cases
of SPR (top) and TPR (bottom). The black point represents the best
fit BPs that corresponds to $BP_{1}:\chi_{(3)}^{2}=3.487$ and $BP_{2}:\chi_{(3)}^{2}=1.796$
for the SPR (top) and TPR cases, respectively. The yellow point in
the upper panels represents the best matching of the excess in the
channels $\gamma\gamma$ and $b\bar{b}$ ($BP_{3}:\chi_{(2)}^{2}\sim10^{-6}$).
In the right panels, the $\chi_{(2)}^{2}$ ($\chi_{(3)}^{2}$) function
(\ref{eq:chi2}) is shown in the upper (lower) palette.}
\label{fig:mu-S}
\end{figure}

Concerning the $95~{\rm GeV}$ signal excess (\ref{eq:excess}), one
defines the functions $\chi_{(N)}^{2}$ as
\begin{equation}
\chi_{(2)}^{2}=\chi_{\gamma\gamma}^{2}+\chi_{b\bar{b}}^{2},~\chi_{(3)}^{2}=\chi_{\gamma\gamma}^{2}+\chi_{b\bar{b}}^{2}+\chi_{\tau\tau}^{2},~\chi_{i}^{2}=\left(\frac{\mu_{i}-\mu_{i}^{\mathrm{exp}}}{\Delta\mu_{i}^{\mathrm{exp}}}\right)^{2},\label{eq:chi2}
\end{equation}
which are useful to check for whether the excess can be addressed simultaneously
in the channels $\gamma\gamma,b\bar{b}$ and/or $\gamma\gamma,\tau\tau,b\bar{b}$,
respectively. In our analysis, we will consider only the BPs that address the three channels simultaneously within
$2-\sigma$, i.e., $\chi_{(3)}^{2}<8.02$. In Fig.~\ref{fig:mu-S},
we show the signal strength modifier values (\ref{eq:mu95}) and the
$\chi_{(2)}^{2}$ ($\chi_{(3)}^{2}$) function for the considered
BPs in the SPR (TPR) scenario in the up (bottom) panels.

From the upper panels in Fig.~\ref{fig:mu-S}, one remarks that the
excess in the channels $\gamma\gamma$ and $b\bar{b}$ is easily matched
simultaneously in the SPR case, where the matching could be exact
for some BPs as $\chi_{(2)}^{2}\sim10^{-6}$. 
According to the minimal value $\chi_{(3)}^{2}=3.487$, the excess could be addressed in the three channels $\gamma\gamma,b\bar{b}$ and $\tau\tau$ simultaneously in $1-\sigma$ for tiny part of the parameter space (0.0059 \% among all BPs) that corresponds to $\chi_{(3)}^{2}<3.53$. This tiny region of the the parameter space exists due to the large value of $\Delta \mu^{\rm exp}_{\tau\tau}=0.5$, so this region could be ruled out once precise measurements are performed for $\mu_{\tau\tau}$ despite the new central value. According to the bottom panels, the $H_{0}^{3}$ contribution to the signal strengths is very
important to address the di-$\tau$ excess in the TPS case. Here,
the $\chi_{(3)}^{2}$ values are getting significantly smaller than
the SPR case. In the TPR case, we have about 27.5\% of the BPs with
less than $1-\sigma$, which means that the excess in the three channels
(\ref{eq:excess}) is addressed simultaneously in the three channels.
Whereas, in the SPR case, the $\chi_{(3)}^{2}$ function values are
larger than $1-\sigma$ for the majority of the BPs considered in Fig.~\ref{fig:mu-S},
as its minimal value is $\chi_{(3)}^{2,min}=3.487$. The best fit
benchmark points $BP_{1}$, $BP_{2}$ and $BP_{3}$ shown in Fig.~\ref{fig:mu-S}
are presented in Table~\ref{T1}.

\begin{table}
\centering%
\begin{tabular}{|c|c|c|c|}
\hline
 & $BP_{1}$ & $BP_{2}$ & $BP_{3}$ \tabularnewline
\hline
\hline
$s_{\beta}$ & -0.511 & -0.496 & 0.345\tabularnewline
\hline
$s_{\alpha}$ & 0.488 & 0.458 & -0.220\tabularnewline
\hline
$m_{3}$ & 94.74 & 103.99 & 100.96\tabularnewline
\hline
$m_{5}$ & 78.15 & 79.38 & 121.62\tabularnewline
\hline
$\kappa_{F}$ & 1.015 & 1.024 & 1.040 \tabularnewline
\hline
$\kappa_{V}$ & 0.903 & 0.911 & 0.962\tabularnewline
\hline
$\kappa_{\gamma}$ & 1.027 & 0.931 & 0.956\tabularnewline
\hline
$\zeta_{F}$ & 0.568 & 0.528 & -0.235\tabularnewline
\hline
$\zeta_{V}$ & -0.308 & -0.322 & 0.344\tabularnewline
\hline
$\zeta_{\gamma}$ & 0.484 & 0.501 & 0.522\tabularnewline
\hline
$\vartheta_{F}$ & -0.489 & -0.458 & 0.220 \tabularnewline
\hline
$\vartheta_{\gamma}$ & 0.149 & 0.140 & 0.067\tabularnewline
\hline
$\mu_{1}$ & -51.31 & -55.48 & 41.42\tabularnewline
\hline
$\mu_{2}$ & -29.69 & -22.17 & 67.95\tabularnewline
\hline
$\mu_{\gamma\gamma}^{(95)}$ & 0.257 & 0.251 & 0.270\tabularnewline
\hline
$\mu_{\tau\tau}^{(95)}$ & 0.562 & 0.278 & 0.054\tabularnewline
\hline
$\mu_{b\bar{b}}^{(95)}$ & 0.095 & 0.104 & 0.117\tabularnewline
\hline
$\rho_{\tau\tau}$ & 0.575 & - & -\tabularnewline
\hline
$\Delta S$ & -0.032 & -0.023 & -0.060 \tabularnewline
\hline
$\chi_{(2)}^{2}$ & 0.168 & 0.088 & $3.7\times10^{-6}$ \tabularnewline
\hline
$\chi_{(3)}^{2}$ & 1.7959 & 3.4868 & 5.2493 \tabularnewline
\hline
$\chi_{SM}^{2}$ & 12.526 & 12.141 & 9.127\tabularnewline
\hline
\end{tabular}\caption{Different physical observables for the BPs shown in Fig.~\ref{fig:mu-S}.
All mass dimension observables are given in GeV. }
\label{T1}
\end{table}

\begin{figure}[h]
\includegraphics[width=0.49\textwidth]{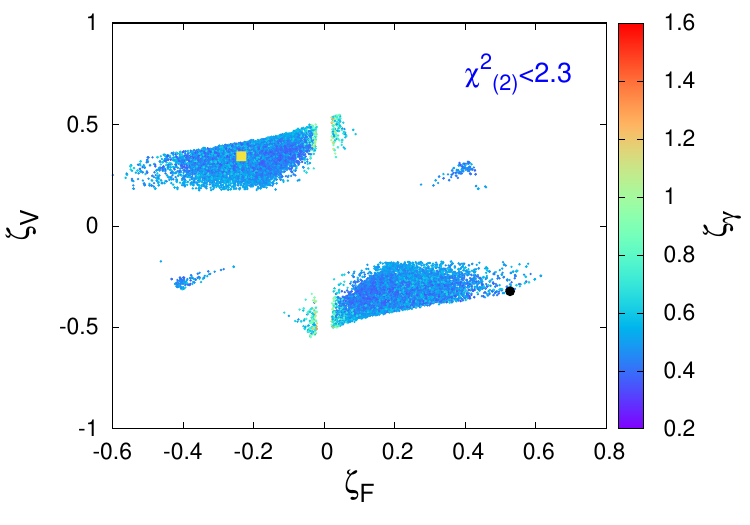}~\includegraphics[width=0.49\textwidth]{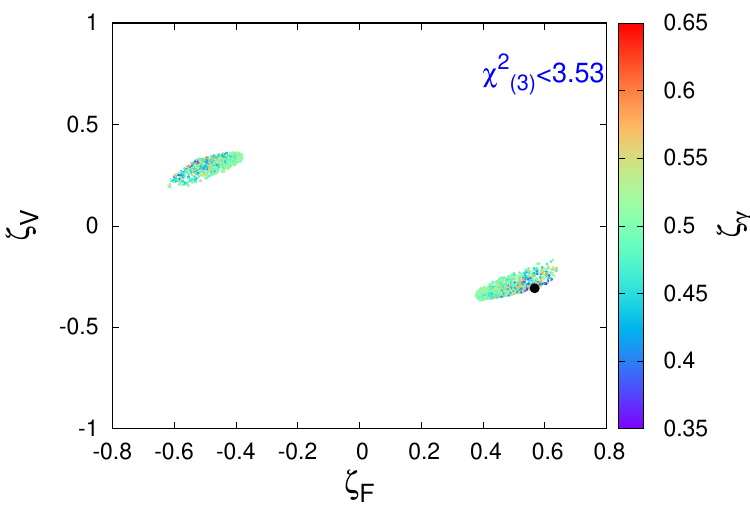}\caption{The couplings modifiers $\zeta_{X}$ for the SPR (right) and TPR (left)
cases. Here, we considered only the $1-\sigma$ BPs, i.e., $\chi_{(2)}^{2}<2.3$
($\chi_{(3)}^{2}<3.53$) for the SPR (TPR) case. The black points
in the middle and right panels correspond to the BPs with $BP_{1}:\chi_{(3)}^{2}=3.487$
and $BP_{2}:\chi_{(3)}^{2}=1.796$, respectively, while the yellow
one represents the BP with $BP_{3}:\chi_{(2)}^{2}=3.57\times10^{-6}$.}
\label{fig:H3}
\end{figure}

For reasons of completeness, we show the couplings modifiers $\zeta_{X}$
for the $1-\sigma$ BPs in the SPR ($\chi_{(2)}^{2}<2.3$) and TPR
($\chi_{(3)}^{2}<3.53$) cases in Fig.~\ref{fig:H3}.

From Fig.~\ref{fig:H3}-right, it is clear that addressing this excess
in the three channels simultaneously makes the parameter space for
the TPR case so tight. However, if the di-$\tau$ excess would be relaxed
to a smaller value like $\mu_{\tau\tau}^{(95)}\sim0.6$ with a good precision $\Delta \mu_{\tau\tau}^{\rm exp}\lesssim 0.1$ in future
analyses, a significant part of the GM parameter space can address
the three measurements simultaneously within both SPR scenario, while for the TPR scenario, we could get an exact matching, .i.e., $\chi^2_{(3)}\sim 0$.

\begin{figure}[h]
\includegraphics[width=0.5\textwidth]{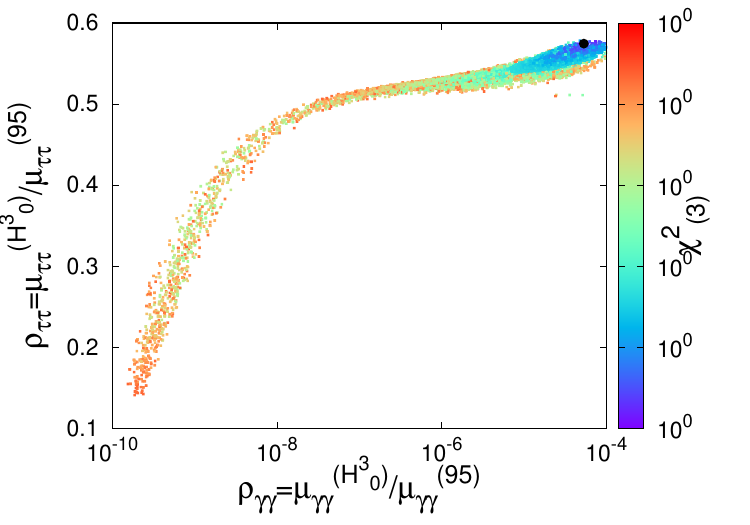}\caption{The ratios $\mu_{\tau\tau}^{(H_{3}^{0})}/\mu_{\tau\tau}^{(95)}$ and
$\mu_{\gamma\gamma}^{(H_{3}^{0})}/\mu_{\gamma\gamma}^{(95)}$ in the
TPR case using the BPs considered in Fig.~\ref{fig:mu-S}. Here,
the palette represents the $\chi_{(3)}^{2}$ function.}
\label{fig:ratio}
\end{figure}

In some attempts to address the excess (\ref{eq:excess}), it is believed
that one of them can be regarded as statistical fluctuations, and
hence, should disappear once more data are collected, for example see~\cite{Aguilar-Saavedra:2023tql}.
If the signal strength $\mu_{\tau\tau}^{(95)}$ will be relaxed to
a smaller value once ATLAS results are reported and/or more data are
considered by CMS, the excess in the three channels can be simultaneously
addressed, even in the SPR case; and, hence the viable parameter space
would be significant. In order to probe the $H_{0}^{3}$ contribution
effect to (\ref{eq:eGam}) in the TPR case, we show the $H_{0}^{3}$
relative contributions to (\ref{eq:eGam}) in Fig.~\ref{fig:ratio}-left.

As expected, the $H_{0}^{3}$ contribution represents $14\sim58\%$
of the signal strength $\mu_{\tau\tau}^{(95)}$, unlike its contribution
to $\mu_{\gamma\gamma}^{(95)}$ that is practically vanishing. From
Fig.~\ref{fig:ratio}, one mentions that large $H_{0}^{3}$ contributions
to $\mu_{\tau\tau}^{(95)}$ are preferred, since they correspond to
smaller values for $\chi_{(3)}^{2}<3.53$.

\section{Distinguishing the SPR and TPR scenarios via the di-$\tau$ Channel\label{sec:Dis}}

If the 95 GeV excess is confirmed in the di-$\tau$ channel when new ATLAS results are reported using more data and/or similar results are released by CMS, this channel could be very useful for distinguishing between the SPR and TPR scenarios. In the SPR case, the CP properties of the 95 GeV signal resonance are well defined, matching those of a CP-even Higgs decaying into $\tau\tau$. However, in the TPR case, the CP properties would be different. Therefore, a mismatch of the CP-even properties in the $\tau\tau$ channel could confirm the TPR scenario.

At the detector level, the $\tau$ lepton cannot be measured directly
but based on its decay products, especially the hadronic final states
${\cal B}(\tau\to had)=64.79\%$~\cite{ParticleDataGroup:2022pth}.
It has two important decay channels $\tau^{\pm}\to\pi^{\pm}\nu_{\tau}$
and $\tau^{\pm}\to\rho^{\pm}\nu_{\tau}\to\pi^{\pm}\pi^{0}\nu_{\tau}$
with the branching ratios 10.82\% and 25.49\%, respectively. However,
the decay $\eta(H_{3}^{0})\to\tau^{+}\tau^{-}\to\pi^{+}\pi^{-}\nu_{\tau}\bar{\nu}_{\tau}$
is more useful to identify the scalar CP properties via the dependence
on the so-called acoplanarity angle that is defined as $\phi^{*}=\arccos(\vec{n}_{+}\cdot\vec{n}_{-})$,
where $\vec{n}_{\pm}$ are unit vectors that are normal to the decay
plans of the charged pions. The $\phi^{*}$ distribution is generally
used to probe the Higgs CP phase $\Delta_{CP}$ of the tau Yukawa
interaction:
\begin{equation}
-\mathcal{L}_{Y}=y_{\tau}h\overline{\psi}_{\tau}(\cos(\Delta_{CP})+i\gamma_{5}\sin(\Delta_{CP}))\psi_{\tau}.\label{eq:LCP}
\end{equation}

The acoplanarity angle normalized distributions for the cases of CP-even,
CP-odd are given by
\begin{align}
R_{even,odd}(\phi^{*}) & =\frac{1}{N}\frac{dN}{d\phi^{*}}=\frac{1}{2\pi}[1\mp Q\cos(\phi^{*})],
\end{align}
with $Q=\frac{\pi^{2}}{16},\frac{\pi^{2}}{16}\Big(\frac{m_{\tau}^{2}-2m_{\rho}^{2}}{m_{\tau}^{2}+2m_{\rho}^{2}}\Big)^{2}$
for the decays $\tau\to\pi^{-}\nu_{\tau}$ and $\tau\to\rho^{-}\nu_{\tau}$,
respectively~\cite{Kuhn:1982di}. While for a degenerate Cp-even/CP-odd
resonance, it is written by
\begin{align}
R_{TPR}(\phi^{*}) & =(1-\rho_{\tau\tau})R_{even}(\phi^{*})+\rho_{\tau\tau}R_{odd}(\phi^{*}),\label{eq:R}
\end{align}
with $\rho_{\tau\tau}=\mu_{\tau\tau}^{(H_{3}^{0})}/\mu_{\tau\tau}^{(95)}$
as presented previously in Fig.~\ref{fig:ratio}. Then, by considering
the maximum/minimum values of the acoplanarity angle distribution
(for example, $\phi^{*}=\pi$ if $\eta$ and $H_{3}^{0}$ are pure
CP-even and CP-odd scalars, respectively), one obtains the ratio $\rho_{\tau\tau}=[1+Q-2\pi\,R_{m}]/[2Q]$,
with $R_{m}$ to be maximum/minimum of $R_{TPR}(\phi^{*})$, which
corresponds to $\phi^{*}=\pi$ for pure CP-even/CP-odd distribution.
This can be easily confirmed numerically.

\begin{figure}[h]
\includegraphics[width=0.5\textwidth]{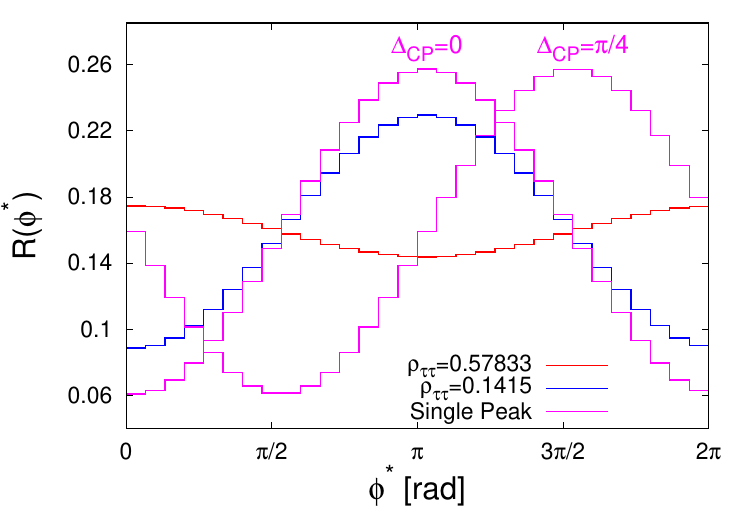}\caption{The acoplanarity angle normalized distributions in the decay $\eta(H_{3}^{0})\to\tau^{+}\tau^{-}\to\pi^{+}\pi^{-}\nu_{\tau}\bar{\nu}_{\tau}$
for two SPR cases ($\Delta_{CP}=0,\pi/4$) in magenta color, and two
TPR BPs with $\rho_{\tau\tau}=0.1415$ (blue) and $\rho_{\tau\tau}=0.57833$
(red).}
\label{fig:tau}
\end{figure}

In Fig.~\ref{fig:tau}, we show the normalized distribution of the
acoplanarity angle ($\phi^{*}$) in the final state $\tau^{+}\tau^{-}\to\pi^{+}\pi^{-}\nu_{\tau}\bar{\nu}_{\tau}$
for the SPR cases of CP conserving ($\Delta_{CP}=0$) and CP violating
($\Delta_{CP}=\pi/4$) (magenta); and two TPR BPs with $\rho_{\tau\tau}=0.1415$
(blue) and $\rho_{\tau\tau}=0.57833$ (red).

One learns from Fig.~\ref{fig:tau} that the TPR case can be easily
identified due to the flatness of the acoplanarity angle distribution;
and can be distinguished from the case of a single resonance, whatever
its CP phase value $\Delta_{CP}$ is for the tau Yukawa interaction in~(\ref{eq:LCP}).
In the case of a pure CP-even (CP-odd) scalar, i.e., $\Delta_{CP}=0$
($\Delta_{CP}=\pi/2$), the $\phi^{*}$ distribution has a maximum
(minimum) at $\phi^{*}=\pi$ with same amplitude. So, obviously, if
the contributions of $\eta$ and $H_{3}^{0}$ are exactly equal, the
distribution (\ref{eq:R}) would be a horizontal straight line. From
Fig.~\ref{fig:tau}, it is clear that the distribution values at
$\phi^{*}=\pi$ for the red and blue curves lead to $\rho_{\tau\tau}=0.1415$
and $\rho_{\tau\tau}=0.57833$, respectively; according to the formula
$\rho_{\tau\tau}=[1+Q-2\pi R_{TPR}(\phi^{*}=\pi)]/[2Q]$ mentioned
previously.

\section{Conclusion\label{sec:Conclusion}}

In this work, we have investigated the possibility of addressing the
$95~{\rm GeV}$ signal excess that is observed in the channels $\gamma\gamma,b\bar{b}$
and $\tau\tau$ within the GM model. We have found that this excess
can be addressed in the GM model in two ways: (1) SPR case where the
signal candidate is a CP-even scalar ($\eta$) with SM-like couplings;
and (2) TPR case, where a CP-odd scalar ($H_{3}^{0}$) whose mass
being around $95~{\rm GeV}$; contributes to this signal excess in
addition to the CP-even scalar $\eta$. By imposing all relevant theoretical
and experimental constraints on the model parameter space, the excess
can be addressed in the channels $\gamma\gamma$ and $b\bar{b}$ simultaneously
in the SPR case ($\chi_{(2)}^{2}\sim10^{-6}$). While in the TPR case,
the three channels $\gamma\gamma$, $b\bar{b}$ and $\tau\tau$ can
be addressed simultaneously ($1.796\leq\chi_{(3)}^{2}\leq3.53$).
This makes the parameter space tight, and the model more predictable
at colliders since the Higgs couplings modifiers are lying in the
ranges $0.90584<\kappa_{F}<1.0591$ and $0.89045<\kappa_{V}<0.99083$
for both SPR and TPR scenarios. Once the ATLAS di-$\tau$ excess will
be reported and/or CMS analysis re-done using more data; and the di-$\tau$
excess may be relaxed to smaller value, let us say around $\mu_{\tau\tau}^{(95)}\sim0.6$ with a good precision $\Delta \mu_{\tau\tau}^{\rm exp}\lesssim 0.1$,
then the GM model will be able to address the excess in the three
channels simultaneous within both SPR and TPR scenarios.

We have shown also that the $95~{\rm GeV}$ candidate scalar di-$\tau$
decay $\eta(H_{3}^{0})\to\tau_{h}^{+}\tau_{h}^{-}$ is very useful
to identify whether it is a SPR or TPR case. The TPR case can be easily
distinguished if the acoplanarity angle normalized distribution gets
flattened with respect to the SPR case. For some specific values of
the acoplanarity angle ($\phi^{*}=\pi$), one can estimate exactly
the relative contributions of both CP-even and CP-odd to the $95~{\rm GeV}$
signal resonance.

\vspace{0.5cm}
\textbf{Acknowledgments}: I would like to thank Salah Nasri for his
useful comments on the manuscript. This work was funded by the University
of Sharjah under the research projects No 21021430107 ``\textit{Hunting
for New Physics at Colliders}'' and No 23021430135 ``\textit{Terascale
Physics: Colliders vs Cosmology}''.


\begin{thebibliography}{1}

\bibitem{ATLAS:2012yve} G.~Aad \textit{et al.} [ATLAS], 
Phys. Lett. B \textbf{716} (2012), 1-29 
 [arXiv:1207.7214 [hep-ex]]. 
S.~Chatrchyan \textit{et al.} [CMS], 
Phys. Lett. B \textbf{716} (2012), 30-61 
[arXiv:1207.7235 [hep-ex]].

\bibitem{ATLAS:2020tlo}G.~Aad \textit{et al.} [ATLAS], 
 Eur. Phys. J. C \textbf{81} (2021) no.4, 332 
 [arXiv:2009.14791 [hep-ex]].

\bibitem{CMS:2018rmh} A.~M.~Sirunyan \textit{et al.} [CMS],
JHEP \textbf{09} (2018), 007 
[arXiv:1803.06553 [hep-ex]].

\bibitem{CMS:2022goy} [CMS], 
[arXiv:2208.02717 [hep-ex]].

\bibitem{ATLAS_talk} C.~Arcangeletti, \emph{ATLAS, LHC Seminar,
\href{https://indico.cern.ch/event/1281604/}{\rm indico.cern.ch/event/1281604}},
2023.

\bibitem{Azatov:2012bz} A.~Azatov, R.~Contino and J.~Galloway,
 JHEP \textbf{04} (2012), 127 [erratum: JHEP \textbf{04} (2013),
140] 
 [arXiv:1202.3415 [hep-ph]].

\bibitem{Cao:2016uwt} J.~Cao, X.~Guo, Y.~He, P.~Wu and Y.~Zhang,
Phys. Rev. D \textbf{95} (2017) no.11, 116001 
[arXiv:1612.08522 [hep-ph]].

\bibitem{Biekotter:2023oen} T.~Biekoetter, S.~Heinemeyer and G.~Weiglein,
 Phys. Rev. D \textbf{109} (2024) no.3, 3 
 [arXiv:2306.03889 [hep-ph]].

\bibitem{Barate:2003sz} R.~Barate \textit{et al.} [LEP Working
Group for Higgs boson searches, ALEPH, DELPHI, L3 and OPAL], 
Phys. Lett. B \textbf{565} (2003), 61-75 
[arXiv:hep-ex/0306033 [hep-ex]].

\bibitem{Cacciapaglia:2016tlr} G.~Cacciapaglia, A.~Deandrea, S.~Gascon-Shotkin,
S.~Le Corre, M.~Lethuillier and J.~Tao, 
JHEP \textbf{12} (2016), 068 
[arXiv:1607.08653 [hep-ph]].

\bibitem{Crivellin:2017upt} A.~Crivellin, J.~Heeck and D.~Moeller,
Phys. Rev. D \textbf{97} (2018) no.3, 035008 
[arXiv:1710.04663 [hep-ph]].

\bibitem{Cao:2019ofo} J.~Cao, X.~Jia, Y.~Yue, H.~Zhou and P.~Zhu,
Phys. Rev. D \textbf{101} (2020) no.5, 055008 
[arXiv:1908.07206 [hep-ph]].

\bibitem{Biekotter:2019kde} T.~Biekoetter, M.~Chakraborti and S.~Heinemeyer,
Eur. Phys. J. C \textbf{80} (2020) no.1, 2 
[arXiv:1903.11661 [hep-ph]].

\bibitem{Cline:2019okt} J.~M.~Cline and T.~Toma, 
Phys. Rev. D \textbf{100} (2019) no.3, 035023 
[arXiv:1906.02175 [hep-ph]].

\bibitem{Abdelalim:2020xfk} A.~A.~Abdelalim, B.~Das, S.~Khalil
and S.~Moretti, 
Nucl. Phys. B \textbf{985} (2022), 116013 
[arXiv:2012.04952 [hep-ph]].

\bibitem{Heinemeyer:2021msz} S.~Heinemeyer, C.~Li, F.~Lika, G.~Moortgat-Pick
and S.~Paasch, 
Phys. Rev. D \textbf{106} (2022) no.7, 075003 
[arXiv:2112.11958 [hep-ph]].

\bibitem{Biekotter:2021qbc} T.~Biekoetter, A.~Grohsjean, S.~Heinemeyer,
C.~Schwanenberger and G.~Weiglein, 
Eur. Phys. J. C \textbf{82} (2022) no.2, 178 
[arXiv:2109.01128 [hep-ph]].

\bibitem{Biekotter:2021ovi} T.~Biekoetter and M.~O.~Olea-Romacho,
JHEP \textbf{10} (2021), 215 
[arXiv:2108.10864 [hep-ph]].

\bibitem{Li:2022etb} W.~Li, J.~Zhu, K.~Wang, S.~Ma, P.~Tian
and H.~Qiao, 
 Chin. Phys. C \textbf{47} (2023) no.12, 123102 
 [arXiv:2212.11739 [hep-ph]].

\bibitem{Biekotter:2022abc} T.~Biekoetter, S.~Heinemeyer and G.~Weiglein,
Eur. Phys. J. C \textbf{83} (2023) no.5, 450 
[arXiv:2204.05975 [hep-ph]].

\bibitem{Benbrik:2022tlg} R.~Benbrik, M.~Boukidi, S.~Moretti and
S.~Semlali, 
PoS \textbf{ICHEP2022} (2022), 547 
[arXiv:2211.11140 [hep-ph]].

\bibitem{Iguro:2022dok} S.~Iguro, T.~Kitahara and Y.~Omura, 
Eur. Phys. J. C \textbf{82} (2022) no.11, 1053 
[arXiv:2205.03187 [hep-ph]].

\bibitem{Biekotter:2022jyr} T.~Biekoetter, S.~Heinemeyer and G.~Weiglein,
JHEP \textbf{08} (2022), 201 
[arXiv:2203.13180 [hep-ph]].

\bibitem{Benbrik:2022azi} R.~Benbrik, M.~Boukidi, S.~Moretti and
S.~Semlali, 
Phys. Lett. B \textbf{832} (2022), 137245 
[arXiv:2204.07470 [hep-ph]]. 

\bibitem{Biekotter:2023jld} T.~Biekoetter, S.~Heinemeyer and G.~Weiglein,
 Phys. Lett. B \textbf{846} (2023), 138217 
 [arXiv:2303.12018 [hep-ph]].

\bibitem{Azevedo:2023zkg} D.~Azevedo, T.~Biekoetter and P.~M.~Ferreira,
 JHEP \textbf{11} (2023), 017 
 [arXiv:2305.19716 [hep-ph]].

\bibitem{Ahriche:2023hho}
A.~Ahriche, M.~L.~Bellilet, M.~O.~Khojali, M.~Kumar and A.~T.~Mulaudzi,
Phys. Rev. D \textbf{110} (2024) no.1, 015025 
[arXiv:2311.08297 [hep-ph]].

\bibitem{Chen:2023bqr} T.~K.~Chen, C.~W.~Chiang, S.~Heinemeyer
and G.~Weiglein, 
 Phys. Rev. D \textbf{109} (2024) no.7, 075043 
 [arXiv:2312.13239 [hep-ph]].

\bibitem{Dev:2023kzu} P.~S.~B.~Dev, R.~N.~Mohapatra and Y.~Zhang,
 Phys. Lett. B \textbf{849} (2024), 138481 
 [arXiv:2312.17733 [hep-ph]].

\bibitem{Li:2023kbf} W.~Li, H.~Qiao, K.~Wang and J.~Zhu, 
[arXiv:2312.17599 [hep-ph]].

\bibitem{Bhattacharya:2023lmu} S.~Bhattacharya, G.~Coloretti, A.~Crivellin,
S.~E.~Dahbi, Y.~Fang, M.~Kumar and B.~Mellado, 
[arXiv:2306.17209 [hep-ph]]. 

\bibitem{Coloretti:2023wng} G.~Coloretti, A.~Crivellin, S.~Bhattacharya
and B.~Mellado, 
Phys. Rev. D \textbf{108} (2023) no.3, 035026 
[arXiv:2302.07276 [hep-ph]]. 

\bibitem{Ashanujjaman:2023etj} S.~Ashanujjaman, S.~Banik, G.~Coloretti,
A.~Crivellin, B.~Mellado and A.~T. Mulaudzi, 
 Phys. Rev. D \textbf{108} (2023) no.9, L091704 
 [arXiv:2306.15722 [hep-ph]].

\bibitem{Liu:2024cbr} C.~X.~Liu, Y.~Zhou, X.~Y.~Zheng, J.~Ma,
T.~F.~Feng and H.~B.~Zhang, 
Phys. Rev. D \textbf{109} (2024) no.5, 056001 
[arXiv:2402.00727 [hep-ph]]. 

\bibitem{Cao:2024axg} J.~Cao, X.~Jia and J.~Lian, 
[arXiv:2402.15847 [hep-ph]].

\bibitem{Kalinowski:2024uxe} J.~Kalinowski and W.~Kotlarski, 
[arXiv:2403.08720 [hep-ph]].

\bibitem{Ellwanger:2024txc} U.~Ellwanger and C.~Hugonie, 
[arXiv:2403.16884 [hep-ph]].

\bibitem{Arcadi:2023smv} G.~Arcadi, G.~Busoni, D.~Cabo-Almeida
and N.~Krishnan, 
[arXiv:2311.14486 [hep-ph]].

\bibitem{Arhrib:2024wjj} A.~Arhrib, K.~H.~Phan, V.~Tran and T.~C.~Yuan,
[arXiv:2405.03127 [hep-ph]]. 

\bibitem{Benbrik:2024ptw} R.~Benbrik, M.~Boukidi and S.~Moretti,
[arXiv:2405.02899 [hep-ph]]. 

\bibitem{Ayazi:2024fmn} S.~Y.~Ayazi, M.~Hosseini, S.~Paktinat
Mehdiabadi and R.~Rouzbehi, 
[arXiv:2405.01132 [hep-ph]]. 

\bibitem{Ellwanger:2024vvs} U.~Ellwanger, C.~Hugonie, S.~F.~King
and S.~Moretti, 
[arXiv:2404.19338 [hep-ph]]. 

\bibitem{Wang:2024bkg} K.~Wang and J.~Zhu, 
[arXiv:2402.11232 [hep-ph]]. 

\bibitem{Georgi:1985nv} H.~Georgi and M.~Machacek, 
Nucl. Phys. B \textbf{262}, 463-477 (1985). 

\bibitem{Ahriche:2022aoj} A.~Ahriche, 
 Phys. Rev. D \textbf{107} (2023) no.1, 015006, Erratum Phys. Rev.
D 108, 019902 (2023) 
 [arXiv:2212.11579 [hep-ph]].

\bibitem{Chanowitz:1985ug} M.~S.~Chanowitz and M.~Golden, 
Phys. Lett. B \textbf{165}, 105-108 (1985) 

\bibitem{Gunion:1989ci} J.~F.~Gunion, R.~Vega and J.~Wudka, 
Phys. Rev. D \textbf{42}, 1673-1691 (1990) 

\bibitem{Haber:1999zh} H.~E.~Haber and H.~E.~Logan, 
Phys. Rev. D \textbf{62}, 015011 (2000) 
 [arXiv:hep-ph/9909335 [hep-ph]].

\bibitem{Aoki:2007ah} M.~Aoki and S.~Kanemura, 
Phys. Rev. D \textbf{77}, no.9, 095009 (2008) [erratum: Phys. Rev.
D \textbf{89}, no.5, 059902 (2014)] 
 [arXiv:0712.4053 [hep-ph]].

\bibitem{Godfrey:2010qb} S.~Godfrey and K.~Moats, 
Phys. Rev. D \textbf{81}, 075026 (2010) 
 [arXiv:1003.3033 [hep-ph]].

\bibitem{Low:2010jp} I.~Low and J.~Lykken, 
JHEP \textbf{10}, 053 (2010) 
[arXiv:1005.0872 [hep-ph]].

\bibitem{Logan:2010en} H.~E.~Logan and M.~A.~Roy, 
Rev. D \textbf{82}, 115011 (2010) 
 [arXiv:1008.4869 [hep-ph]].

\bibitem{Chang:2012gn} S.~Chang, C.~A.~Newby, N.~Raj and C.~Wanotayaroj,
Phys. Rev. D \textbf{86}, 095015 (2012) 
 [arXiv:1207.0493 [hep-ph]].

\bibitem{Kanemura:2013mc} S.~Kanemura, M.~Kikuchi and K.~Yagyu,
Phys. Rev. D \textbf{88}, 015020 (2013) 
 [arXiv:1301.7303 [hep-ph]].

\bibitem{Englert:2013zpa} C.~Englert, E.~Re and M.~Spannowsky,
Rev. D \textbf{87}, no.9, 095014 (2013) 
 [arXiv:1302.6505 [hep-ph]].

\bibitem{Killick:2013mya} R.~Killick, K.~Kumar and H.~E.~Logan,
033015 (2013) 
[arXiv:1305.7236 [hep-ph]].

\bibitem{Englert:2013wga} C.~Englert, E.~Re and M.~Spannowsky,
035024 (2013) 
[arXiv:1306.6228 [hep-ph]].

\bibitem{Ghosh:2019qie} N.~Ghosh, S.~Ghosh and I.~Saha, 
Phys. Rev. D \textbf{101} (2020) no.1, 015029 
 [arXiv:1908.00396 [hep-ph]].

\bibitem{Das:2018vkv} D.~Das and I.~Saha, 
Phys. Rev. D \textbf{98} (2018) no.9, 095010 
 [arXiv:1811.00979 [hep-ph]].

\bibitem{Hartling:2014zca} K.~Hartling, K.~Kumar and H.~E.~Logan,
Phys. Rev. D \textbf{90} (2014) no.1, 015007 
 [arXiv:1404.2640 [hep-ph]].

\bibitem{Hartling:2014aga} K.~Hartling, K.~Kumar and H.~E.~Logan,
Phys. Rev. D \textbf{91} (2015) no.1, 015013 
[arXiv:1410.5538 [hep-ph]].

\bibitem{Chiang:2014bia} C.~W.~Chiang, S.~Kanemura and K.~Yagyu,
Phys. Rev. D \textbf{90} (2014) no.11, 115025 
[arXiv:1407.5053 [hep-ph]].

\bibitem{Chiang:2015rva} C.~W.~Chiang, S.~Kanemura and K.~Yagyu,
Phys. Rev. D \textbf{93} (2016) no.5, 055002 
[arXiv:1510.06297 [hep-ph]].

\bibitem{Chang:2017niy} J.~Chang, C.~R.~Chen and C.~W.~Chiang,
JHEP \textbf{03} (2017), 137 
[arXiv:1701.06291 [hep-ph]].

\bibitem{Chiang:2015kka} C.~W.~Chiang and K.~Tsumura, 
JHEP \textbf{04} (2015), 113 
[arXiv:1501.04257 [hep-ph]].

\bibitem{Chen:2022zsh} T.~K.~Chen, C.~W.~Chiang, C.~T.~Huang
and B.~Q.~Lu, 
[arXiv:2205.02064 [hep-ph]].

\bibitem{Chen:2020ark} S.~L.~Chen, A.~Dutta Banik and Z.~K.~Liu,
Nucl. Phys. B \textbf{966} (2021), 115394 
[arXiv:2011.13551 [hep-ph]].

\bibitem{Pilkington:2017qam} T.~Pilkington, 
[arXiv:1711.04378 [hep-ph]].

\bibitem{Chiang:2014hia} C.~W.~Chiang and T.~Yamada, 
Phys. Lett. B \textbf{735} (2014), 295-300 
[arXiv:1404.5182 [hep-ph]].

\bibitem{Ismail:2020zoz} A.~Ismail, H.~E.~Logan and Y.~Wu, 
 [arXiv:2003.02272 [hep-ph]].

\bibitem{Bairi:2022adc} Z.~Bairi and A.~Ahriche, 
Phys. Rev. D \textbf{108} (2023) no.5, 5 
[arXiv:2207.00142 [hep-ph]]. 

\bibitem{Ghosh:2023izq} S.~Ghosh, 
[arXiv:2311.15405 [hep-ph]]. 

\bibitem{Djouadi:2005gi} A.~Djouadi, 
Phys. Rept. \textbf{457} (2008), 1-216 
[arXiv:hep-ph/0503172 [hep-ph]].

\bibitem{Higgs} The LHC Higgs Working Group: 
\url{https://twiki.cern.ch/twiki/bin/view/LHCPhysics/LHCHWG}

\bibitem{ParticleDataGroup:2022pth} R.~L.~Workman \textit{et al.}
[Particle Data Group], 
PTEP \textbf{2022} (2022), 083C01. 

\bibitem{Gunion:1990dt} J.~F.~Gunion, R.~Vega and J.~Wudka, 
 Phys. Rev. D \textbf{43} (1991), 2322-2336. 

\bibitem{CMS:2017fhs} A.~M.~Sirunyan \textit{et al.} [CMS],
 Phys. Rev. Lett. \textbf{120} (2018) no.8, 081801 
[arXiv:1709.05822 [hep-ex]].

\bibitem{ATLAS:2014jdv} G.~Aad \textit{et al.} [ATLAS], 
 Phys. Rev. Lett. \textbf{113} (2014) no.17, 171801 
 [arXiv:1407.6583 [hep-ex]].

\bibitem{ATLAS:2017ayi} M.~Aaboud \textit{et al.} [ATLAS], 
 Phys. Lett. B \textbf{775} (2017), 105-125 
 [arXiv:1707.04147 [hep-ex]].

\bibitem{OPAL:2002ifx} G.~Abbiendi \textit{et al.} [OPAL], 
Eur. Phys. J. C \textbf{27} (2003), 311-329 
[arXiv:hep-ex/0206022 [hep-ex]].

\bibitem{CMS:2018cyk} A.~M.~Sirunyan \textit{et al.} [CMS],
Phys. Lett. B \textbf{793} (2019), 320-347 
[arXiv:1811.08459 [hep-ex]].

\bibitem{Misiak:2006zs} M.~Misiak, H.~M.~Asatrian, K.~Bieri,
M.~Czakon, A.~Czarnecki, T.~Ewerth, A.~Ferroglia, P.~Gambino,
M.~Gorbahn and C.~Greub, \textit{et al.} 
Phys. Rev. Lett. \textbf{98} (2007), 022002 
[arXiv:hep-ph/0609232 [hep-ph]].

\bibitem{Becher:2006pu}
T.~Becher and M.~Neubert,
Phys. Rev. Lett. \textbf{98} (2007), 022003
[arXiv:hep-ph/0610067 [hep-ph]].

\bibitem{Aguilar-Saavedra:2023tql} J.~A.~Aguilar-Saavedra, H.~B.~Camara,
F.~R.~Joaquim and J.~F.~Seabra, 
Phys. Rev. D \textbf{108} (2023) no.7, 075020 
[arXiv:2307.03768 [hep-ph]]. 

\bibitem{Kuhn:1982di} J.~H.~Kuhn and F.~Wagner, 
Nucl. Phys. B \textbf{236} (1984), 16-34 

\end{thebibliography}
\end{document}